\pdfoutput=1
\documentclass[11pt]{article}
\usepackage{graphicx}
\usepackage{slashbox}
\usepackage{multirow}
\usepackage{bm}

\catcode`@=12
\makeatletter
\@addtoreset{equation}{section}

\makeatother
\begin{document}
\title{\LARGE \bf Scaling ansatz with texture zeros in linear seesaw }
\author{{\bf  Mainak Chakraborty$^{\rm a}$\footnote{mainak.chakraborty@saha.ac.in}, H. Zeen Devi$^{\rm b}$\footnote{zdevi@utm.ac.in},
Ambar Ghosal$^{\rm a}$\footnote{ambar.ghosal@saha.ac.in}}\\
 a) Saha Institute of Nuclear Physics, 1/AF Bidhannagar,
  Kolkata 700064, India \\
b) University of Technology and Management, Shillong 793003, India\hspace{1cm}
 }
%\date{}
\maketitle
%\keywords{Neutrino Physics, Beyond Standard Model}
%\email{ambar.ghosal@saha.ac.in}
\begin{abstract}
 We investigate scaling ansatz with texture zeros within the framework of linear seesaw mechanism. In this variant of seesaw
mechanism a simplified expression of effective neutrino mass matrix $m_\nu$ containing two Dirac type matrices ($m_D$ and 
$m_{DS}$) and one Majorana type matrix ($m_{RS}$) is obtained by virtue of neglecting the global $U(1)_L$ symmetry breaking 
term in the mass term of the Lagrangian. Along with the charged lepton mass matrix, the matrix $m_{RS}$ too, is chosen in a 
diagonal basis whereas a scaling relation is incorporated in $m_D$ and $m_{DS}$ with different scale factors. Our goal in 
this work is to achieve a completely phenomenologically acceptable $m_\nu$ generated by combinations of $m_D$ and $m_{DS}$ 
containing least number of independent parameters or maximum number of zeros. At the end of the numerical analysis it is found
that number of zeros in any of the constituent Dirac type matrices ($m_D$ and $m_{DS}$) of $m_\nu$ cannot be greater than six 
in order to meet the phenomenological requirements. The hierarchy obtained here is normal and also the values of the two 
parameters sum mass ($\sum m_i$) and $|m_{\nu_{ee}}|$ are below the present experimental lower limit.
\end{abstract}
\newpage
\section{Introduction}
In the quest towards understanding of a viable flavour structure of low-energy neutrino mass matrix adhering neutrino oscillation data,
a general approach is to advocate flavour symmetries in conjunction with the standard $SU(2)_L\times U(1)_Y$ model. Those additional
flavour symmetries are associated with some gauge group, discrete or continuous, and thereby dictating a well-defined theory
to explain the extant data. This is a task to realize a comprehensive theory in the ultimate goal to comply with all experimental 
results. On the contrary, realization of a viable neutrino mass matrix through the proposition of some ansatz at low energy is also
a supportive way towards the quest of a more elucidative model.
\paragraph{}
In the present work we investigate the latter idea considering two ansatzes, (i)zeros in the Yukawa texture, (ii)a scaling property
between the nonzero Yukawa matrix elements, referred to as scaling ansatz\cite{sc1}-\cite{Adhikary:2012kb} within the framework of a variant of seesaw mechanism 
known as \textquotedblleft linear seesaw\textquotedblright mechanism\cite{Hettmansperger:2011bt}-\cite{Hirsch:2009mx}. We do not touch the origin of those two well-studied 
ansatzes here, however, we bring the two ansatzs together here and investigate systematically the minimal number of parameters
necessary to explain the neutrino experimental data within the above mentioned framework. We briefly mention few words regarding
scaling ansatz. Imposition of scaling ansatz correlates the nonzero elements of Yukawa matrix by a scale factor and it can be 
achieved through different ways. One of the distinctive properties of scaling ansatz is that the texture remains invariant under 
renormalization group evolution. Furthermore, this ansatz leads to $m_3=0$ and $\theta_{13}=0$. Thus we are compelled to 
break the ansatz to generate nonzero $\theta_{13}$.
\paragraph{}
Texture zeros\cite{Frampton:2002yf}-\cite{4zero4} are investigated in the literature within different framework to generate light neutrino masses. Here, we start with
maximum number of zeros in Yukawa matrix and investigate by reducing the number of zeros till we get a minimum of
necessary parameters to explain neutrino oscillation data\cite{Forero:2014bxa, GonzalezGarcia:2012sz, Tortola:2012te}.
\paragraph{}
Our plan of the paper is as follows. Section \ref{s2} deals with linear seesaw mechanism framework. The scaling ansatz considered
is given in section \ref{s3}. Section \ref{s4} contains analysis with texture zeros. Parametrization and diagonalization of the 
emerged neutrino mass matrices is shown in Section \ref{s5}. 
Discussion on numerical result is given in section \ref{s6}.
Section \ref{s7} contains the summary and conclusion of the present work.
%Section \label{s6} contains discussion on numerical results.
%Summary of the present work is given in Section \ref{s7}.  
\section{Linear seesaw}\label{s2}
%In most of the neutrino mass models the smallness of neutrino mass is achieved through seesaw mechanism. 
In linear seesaw, the effective neutrino mass matrix ($m_\nu$) generated varies linearly with Dirac neutrino mass matrix ($m_D$)
instead of quadratic variation as happens in type-I seesaw.
In this popular variant of type-I seesaw along with left chiral SM doublet neutrinos ($\nu_L$) and right chiral singlet neutrinos ($N_{iR}$),
extra fermion singlets ($S_{iR}$) are added. In effect the well-known type-I seesaw  basis $((\nu^c)_R,N_R)$ is extended to $((\nu^c)_R,N_R,S_R)$. 
Linear seesaw mechanism arises when the mass matrix in the above basis takes the following form
\begin{equation}
M_\nu=\left( \begin{array}{ccc}
     0 & m_D & m_{DS}\\m_D^T & 0 & m_{RS}\\m_{DS}^T& m_{RS}^T& M_S\\ 
    \end{array}\right)
\label{mnu1}
\end{equation}
where $M_\nu$ is a $9\times 9$ matrix assuming three generations of each fermions. 
%since we have considered 3 generations of each of the fermions ($\nu_L$, $N_R$, $S$).
%We have to block diagonalize the above matrix to get the light neutrino mass matrix. For this purpose two new matrices
To obtain light neutrino masses, we have to block diagonalize (\ref{mnu1}) and with the introduction of the following matrices
\begin{eqnarray}
M_D=\left( \begin{array}{cc}
      m_D & m_{DS}\\ 
    \end{array}\right),\nonumber\\
M_R=\left( \begin{array}{cc}
      0 & m_{RS}\\m_{RS}^T & M_S\\
    \end{array}\right)
\label{block}
\end{eqnarray}
%are introduced which upon substitution in eq.(\ref{mnu1}) gives
the effective $M_\nu$ takes the form as
\begin{equation}
M_\nu=\left( \begin{array}{cc}
      0 & M_D\\M_D^T & M_R\\
    \end{array}\right)
\label{mnu2}
\end{equation}
which is exactly similar to that of type-I seesaw mass matrix. Further assuming the hierarchy of type-I seesaw mechanism
the light neutrino mass matrix is obtained easily as
\begin{eqnarray}
m_\nu&=&-M_D M_R^{-1} M_D^T\nonumber\\
     &=&m_D(m_{RS}^T)^{-1}M_S(m_{RS})^{-1}m_{DS}^T-m_D(m_{RS}^T)^{-1}m_{DS}^T-m_{DS}m_{RS}^{-1}m_{D}^T.\label{mn}
\end{eqnarray}
If the $U(1)_L$ global lepton number symmetry breaking term $M_S$ is absent, 
(\ref{mn}) is then simply reduced to 
\begin{equation}
m_\nu=-m_D(m_{RS}^T)^{-1}m_{DS}^T-m_{DS}m_{RS}^{-1}m_{D}^T.\label{mnu4}
\end{equation}
This is our main working formula for the present work and we proceed further to calculate light neutrino masses
and mixing angles with this $m_\nu$ imposing scaling ansatz and texture zeros on the mass matrices $m_D$ and $m_{DS}$.
Moreover, without any loss of generality, we assume that the charged lepton mass matrix and $m_{RS}$ are flavour diagonal.
With such choice of basis it is not possible to consider further 
any other matrix 
flavour diagonal. 
\section{Scaling ansatz}\label{s3}
There are several works\cite{sc1}-\cite{sc3} in which the scaling ansatz has been studied through its imposition on the columns of $m_\nu$.
In the present work we consider scaling ansatz at a more fundamental level through its implementation in $m_D$ and $m_{DS}$.
Furthermore, we impose this ansatz along the rows of the $m_D$ and $m_{DS}$ matrices and we find that such choice of
$m_D$ and $m_{DS}$ leads to the same structure of $m_\nu$, after invoking linear seesaw mechanism. 
% which are mixing terms of $\nu_L$ with $N_R$ and $\nu_L$ with $S$ respectively. 
Scaling ansatz dictates that all the elements of a certain
row of $m_D$ (or $m_{DS}$ ) are related to the elements of another row by a definite scale factor. This scaling relation
may be of three types as (i) the first and second row are connected, 
(ii) the second and third row are connected or (iii) the first and third
row are connected. The cases (i) and (iii) lead to $\theta_{12}$ or $\theta_{23}$ equals to zero and hence, we discard those cases. 
Here we carry out our analysis for the case (ii) which is explicitly written as
\begin{eqnarray}
&&{(m_D)}_{\mu i} =k{(m_D)}_{\tau i} \nonumber\\
&&{(m_{DS})}_{\mu i} =k_1{(m_{DS})}_{\tau i} ,
\label{scale} 
\end{eqnarray}
where $i$ is the column index ($i=1,2,3$) and $k$, $k_1$ are the scale factors for $m_D$ and $m_{DS}$ respectively.   
We now check the effect of the scaling ansatz in the effective neutrino mass matrix $m_\nu$. Using the linear seesaw
formula (\ref{mnu4}) we obtain $m_\nu$ as
\begin{eqnarray}
(m_\nu)_{\mu \alpha}&=& -[(m_D)_{\mu j}(m_{RS}^T)^{-1}_{j l}(m_{DS})^T_{l \alpha}+(m_{DS})_{\mu j}(m_{RS})^{-1}_{j l}(m_{D})^T_{l \alpha}]\nonumber\\
                    &=&-[k(m_D)_{\tau j}(m_{RS}^T)^{-1}_{j l}(m_{DS})^T_{l \alpha}+k_1(m_{DS})_{\tau j}(m_{RS})^{-1}_{j l}(m_{D})^T_{l \alpha}]
\label{mnu5}
\end{eqnarray}
where sum over repeated index is implied. It is clear from the above equation that the scaling ansatz is already broken by the 
choice of different scale factors for $m_D$ and $m_{DS}$. The ansatz can be restored in $m_\nu$ simply by choosing $k=k_1$ and
then (\ref{mnu5}) becomes
\begin{equation}
(m_\nu)_{\mu \alpha}=k(m_\nu)_{\tau \alpha} 
\end{equation}
with $\alpha=e,~\mu,~\tau$ and the scaling relations in $m_\nu$ are obtained as
\begin{equation}
\frac{ (m_\nu)_{\mu e}}{(m_\nu)_{\tau e}}=\frac{(m_\nu)_{\mu\mu}}{(m_\nu)_{\tau\mu}}=
\frac{(m_\nu)_{\mu\tau}}{(m_\nu)_{\tau\tau}} = k .
\label{a1}\end{equation}
As we are aware of the fact that such type of scaling ansatz invariant matrices yield $\theta_{13}=0$, we are compelled to 
deal with the case where $k \neq k_1$. The explicit forms of $m_D$ and $m_{DS}$ with scaling ansatz are given by
\begin{eqnarray}
&&m_D=\left( \begin{array}{ccc}
      a_1 & a_2 & a_3\\ kb_1 & kb_2  & kb_3\\b_1 &b_2 &b_3 \\
    \end{array}\right),\nonumber\\
&&m_{DS}=\left( \begin{array}{ccc}
      x_1 & x_2 & x_3\\ k_1y_1 & k_1y_2  & k_1y_3\\y_1 &y_2 &y_3 \\
    \end{array}\right)
\end{eqnarray}
and as mentioned earlier $m_{RS}$ is taken diagonal as
\begin{equation}
m_{RS}=diag(m_1,~m_2,~m_3) .
\end{equation}
\section{Texture zeros}\label{s4}
In our scheme developed we further put constraint on $m_D$ and $m_{DS}$ through the imposition of zeros and 
our aim here is to find out the maximum number of zeros that we can accommodate in $m_D$ and $m_{DS}$ which 
will produce a phenomenologically viable $m_\nu$. We start our analysis with 8 zero texture, and then move on by reducing 
the number of zeros. We calculate $m_\nu$ for all possible combinations of $m_D$ and $m_{DS}$ and check how many of them can 
give rise to nonzero mixing angles and mass squared differences and how many can be ruled out at once using suitable arguments.
%\paragraph{}
First we tabulate scaling ansatz invariant $n$ zero (where $n=8,7,6,..$) textures of $m_D$ (and $m_{DS}$).
In Table \ref{t1} we represent the scaling ansatz invariant texture zero structures of $m_D$ matrices. 
First of all $m_D$ with $8$ zeros is not possible because scaling ansatz requires at least two nonzero elements 
(one in each row connected by scaling). Texture with $7$ zeros and $6$ zeros are allowed due to compatibility with 
scaling ansatz.  
\begin{table}[!h]
\caption{List of $n$ zero $m_D$ matrices ($n=8,7,6$)} \label{t1}
\begin{tabular}{|c|c|c|}
\hline
\multicolumn{3}{|c|}{{\bf $8$ zero texture}}\\
\hline
\multicolumn{3}{|c|}{ No allowed texture}\\
\multicolumn{3}{|c|}{ }\\
\hline 
\multicolumn{3}{|c|}{{\bf $7$ zero texture}}\\
\hline
$a_1=a_2=a_3=0$ & $a_1=a_2=a_3=0$ & $a_1=a_2=a_3=0$ \\
and $b_2=b_3=0$&  and $b_1=b_3=0$ &  and $b_1=b_2=0$ \\
$m_D(1)=\left(\begin{array}{ccc} 0 & 0 & 0\\ 
                k b_1 & 0 & 0\\
                b_1 & 0  & 0  \end{array}\right) $
& $m_D(2)= \left(\begin{array}{ccc}0 & 0 & 0\\ 
                0 & kb_{2} & 0\\
                0 & b_{2}  & 0 \end{array}\right)$
& $m_D(3)=\left(\begin{array}{ccc} 0 & 0 & 0\\ 
                0 & 0 & k b_3\\
                0 & 0  & b_3 \end{array}\right)$ \\
\hline
\multicolumn{3}{|c|}{{\bf $6$ zero texture}}\\
\hline
$a_{2}=a_{3}=0$ & $a_{1}=a_{3}=0$ & $a_{1}=a_{2}=0$\\
and $b_{2}=b_{3}=0$ & and $b_{2}=b_{3}=0$&and $b_{2}=b_{3}=0$ \\
 $m_D(1)=\left(\begin{array}{ccc} a_{1} & 0 &0\\ 
             k b_1 & 0 & 0\\
             b1 & 0  & 0 \end{array}\right)$
& $m_D(2)=\left(\begin{array}{ccc} 0 & a_{2} &0\\ 
             kb_{1} & 0 & 0\\
             b_{1} & 0  & 0 \end{array}\right)$
&  $m_D(3)=\left(\begin{array}{ccc} 0 & 0 & a_{3}\\ 
             kb_{1} & 0 & 0\\
             b_{1} & 0  & 0 \end{array}\right)$ \\
\hline 
$a_{1}=a_{3}=0$ & $a_{2}=a_{3}=0$ & $a_{1}=a_{2}=0$\\
and $b_{1}=b_{3}=0$ & and $b_{1}=b_{3}=0$&and $b_{1}=b_{3}=0$ \\
 $m_D(4)=\left(\begin{array}{ccc} 0 & a_2 &0\\ 
             0 & k b_2 & 0\\
             0 & b_2  & 0 \end{array}\right)$
& $m_D(5)=\left(\begin{array}{ccc} a_1 & 0 &0\\ 
             0 & kb_2 & 0\\
             0 & b_2  & 0 \end{array}\right)$
&  $m_D(6)=\left(\begin{array}{ccc} 0 & 0 & a_{3}\\ 
             0 & k b_2 & 0\\
             0 & b_2  & 0 \end{array}\right)$ \\
\hline 
$a_{1}=a_{2}=0$ & $a_{2}=a_{3}=0$ & $a_{1}=a_{3}=0$\\
and $b_{1}=b_{2}=0$ & and $b_{1}=b_{2}=0$&and $b_{1}=b_{2}=0$ \\
 $m_D(7)=\left(\begin{array}{ccc} 0 & 0 & a_3\\ 
             0 & 0 & k b_3\\
             0 & 0  & b_3 \end{array}\right)$
& $m_D(8)=\left(\begin{array}{ccc} a1 & 0 &0\\ 
             0 & 0 & k b_3\\
             0 & 0  & b_3 \end{array}\right)$
&  $m_D(9)=\left(\begin{array}{ccc} 0 & a_2 & 0\\ 
             0 & 0 & k b_3\\
             0 & 0  & b_3 \end{array}\right)$ \\
\hline 
\end{tabular}
\end{table}   
A table completely identical to Table \ref{t1} can be constructed for $m_{DS}$ matrix simply by the following substitutions:
$a_i \rightarrow x_i$, $b_i \rightarrow y_i$ and $k \rightarrow k_1$ (where $i=1,2,3$). Thus there are three $7$ zero and 
nine $6$ zero textures allowed for both $m_D$ and $m_{DS}$.
Now we calculate $m_\nu$ using linear seesaw formula (\ref{mnu5}) for all possible combinations of $m_D$ and $m_{DS}$ and 
there are altogether $144$ different possible combinations.
%tabulate them in a convenient way.
Depending upon the position of zeros in the resulting $m_\nu$ matrices we divide those $144$ textures
in $8$ classes and denote them as
\begin{eqnarray}
&&t_1=\left(\begin{array}{ccc} \times & \times & \times\\ 
             \times & \times & \times\\
             \times & \times  & \times \end{array}\right),~
t_2=\left(\begin{array}{ccc} 0& \times & \times\\ 
             \times & \times & \times\\
             \times & \times  & \times \end{array}\right),~
t_3=\left(\begin{array}{ccc} \times & 0 & 0\\ 
             0 & \times & \times\\
             0 & \times  & \times \end{array}\right),~
t_4=\left(\begin{array}{ccc} \times & \times & \times\\ 
             \times & 0 & 0\\
             \times & 0  & 0 \end{array}\right),\nonumber\\
&&t_5=\left(\begin{array}{ccc} 0 & 0 & 0\\ 
             0 & \times & \times\\
             0 & \times  & \times \end{array}\right),~
t_6=\left(\begin{array}{ccc} 0& \times & \times\\ 
             \times & 0 & 0\\
             \times & 0  & 0 \end{array}\right),~
t_7=\left(\begin{array}{ccc} \times & 0 & 0\\ 
             0 & 0 & 0\\
             0 & 0  & 0 \end{array}\right),~
t_8={\rm{null~matrix}} \label{ts}
%t_8=\left(\begin{array}{ccc} 0 & 0 & 0\\ 
%             0 & 0 & 0\\
%             0 & 0  & 0 \end{array}\right)\label{ts}
\end{eqnarray}
where $\times$ denotes generically some nonzero element. Exact expression of $\times$ comes out from the structure of
corresponding $m_D$ and $m_{DS}$. Explicitly emergence of all those classified forms are shown in Tables \ref{t2}-\ref{t4}.
The tables of $m_\nu$ (Tables \ref{t2}-\ref{t4}) are presented in a matrix form in which along the row we write the index of $m_D$ (denoted as $i$)
and along the columns the index of $m_{DS}$ (denoted as $j$) is assigned. Hence the $ij$th element of the table denotes
the type of $m_\nu$ generated by the combination of $m_{D}(i)$ and $m_{DS}(j)$.
We mark the surviving textures by bold letters in those tables. 
%we also encircled in the table the surviving
%textures which we discuss in the following.
\paragraph{}
Lets first check how many of these $t_i$ matrices have the potential to generate phenomenologically
viable mixing angles and mass eigenvalues. It has been shown by Frampton et.al\cite{Frampton:2002yf} that if the number of independent
zeros in an effective neutrino mass matrix ($m_\nu$) is $\geq$ 3, that matrix doesn't favour the oscillation data. This 
result drastically eliminates the matrices from $t_4$ to $t_8$.
Thus the matrices given in given in Table \ref{t2} (containing $7$ zero $m_D$ and $7$ zero $m_{DS}$) are all ruled out.
Again although $t_3$ matrix survives this criteria ( since the number of independent zeros in $t_3$ matrix is only $2$), 
however, one generation
%The structure of the matrix $t_3$ is such that one generation
of neutrino is completely decoupled from the other two, and as a result two mixing angles become zero. Hence we also neglect $t_3$ matrix.
Thus the number of surviving $t_i$ matrix is only $2$ and they are $t_1$ and $t_2$. The matrix $t_1$ appears only in Table \ref{t5}
due to three different combinations of $m_D$ and $m_{DS}$ ($6$ zero $m_D$ with $6$ zero $m_{DS}$) and matrix $t_2$ 
appears in Tables \ref{t3}-\ref{t5} due to total $18$ different combinations of the above matrices. Both Tables \ref{t3}
(combination of $6$ zero $m_D$ and $7$ zero $m_{DS}$) and \ref{t4} ($7$ zero $m_D$ and $6$ zero $m_{DS}$) give three $t_2$
matrices each and Table \ref{t5} gives three $t_1$ matrices and twelve $t_2$ matrices.
%So $t_3$ can also be neglected. We are left with only $t_1$ and $t_2$ which are acceptable for further analysis.
%\paragraph{}
%The tables for $m_\nu$ are constructed in a matrix form in which along the row we write the index of $m_D$ (denoted as $i$)
%and the index of $m_{DS}$ (denoted as $j$) is assigned along the column. So the $ij$th element of the table denotes
%the type of $m_\nu$ generated by the combination of $m_{D}(i)$ and $m_{DS}(j)$. (We mark the surviving textures by
%bold letters in those tables.)
%%%%%%%%%%%%%%%%%% 7,0 md & 7,0 mds %%%%%%%%%%%%%%%%%%%%%%%%%%%%%%%%
\begin{table}[!h]
\caption{Combination of $7$ zero $m_D(i)$ and $7$ zero $m_{DS}(j)$} \label{t2}
\begin{center}
\begin{tabular}{p{2cm}|p{1cm}p{1cm}p{1cm}|}
%\begin{tabular}{c|c|c|}
%\hline\\
\cline{2-4}
 & \multicolumn{3}{c|}{\bf Type of $m_\nu$}\\%{{\bf $8$ zero texture}}\\
\hline
\multicolumn{1}{ |c| }{\backslashbox{ $j$}{ $i$}} &  1 & 2 & 3 \\ \cline{1-4} %\hline
\multicolumn{1}{ |c| }{1 } &  $t_5$ & $t_8$ & $t_8$\\
\multicolumn{1}{ |c| }{2 } & $t_8$ & $t_5$ & $t_8$\\
\multicolumn{1}{ |c| }{3 } & $t_8$ & $t_8$ & $t_5$\\
\hline
\end{tabular}
\end{center}
\end{table}
%%%%%%%%%%%%%%%%%%%%%%%%%%%%%%%%%%%%%%%%%%%%%%%%%%%%%%%%%%%%%%%

%%%%%%%%% 6 zero md & 7 zero mds%%%%%%%%%%%%%%%%%%%%%%%%%
\begin{table}[!h]
\caption{Combination of $6$ zero $m_D(i)$ and $7$ zero $m_{DS}(j)$} \label{t3}
\begin{center}
\begin{tabular}{p{2cm}|p{.6cm}p{.6cm}p{.6cm}p{.6cm}p{.6cm}p{.6cm}p{.6cm}p{.6cm}p{.6cm}|}
%\begin{tabular}{c|c|c|}
%\hline\\
\cline{2-10}
 & \multicolumn{9}{c|}{\bf Type of $m_\nu$}\\%{{\bf $8$ zero texture}}\\
\hline
\multicolumn{1}{ |c| }{\backslashbox{$j$}{$i$}} &  1 & 2 & 3 & 4& 5& 6 & 7 & 8 & 9\\ \cline{1-10} %\hline
\multicolumn{1}{ |c| }{1 } & ${\bm t_2}$  & $t_5$ & $t_5$ & $t_8$ & $t_6$ & $t_8$ & $t_8$ & $t_6$ & $t_8$\\
\multicolumn{1}{ |c| }{2 } & $t_8$ & $t_6$ & $t_8$ & ${\bm t_2}$ & $t_5$ & $t_6$ & $t_8$ & $t_8$ & $t_6$ \\
\multicolumn{1}{ |c| }{3 } & $t_8$ & $t_8$ & $t_6$ & $t_8$&$t_8$ &$t_6$ & ${\bm t_2}$ & $t_5$&$t_5$\\
\hline
\end{tabular}
\end{center}
\end{table}
%%%%%%%%%%%%%%%%%%%%%%%%%%%%%%%%%%%%%%%%%%%%%%%%%%%%%%%%%%%%%

%%%%%%%%%%%%%%%%%%%% $7$ zero $m_D$ and $6$ zero $m_{DS}$%%%%%%%%%%%%%%%%%
\begin{table}[!h]
\caption{Combination of $7$ zero $m_D(i)$ and $6$ zero $m_{DS}(j)$} \label{t4}
\begin{center}
\begin{tabular}{p{2cm}|p{1cm}p{1cm}p{1cm}|}
%\begin{tabular}{c|c|c|}
%\hline\\
\cline{2-4}
 & \multicolumn{3}{c|}{\bf Type of $m_\nu$}\\%{{\bf $8$ zero texture}}\\
\hline
\multicolumn{1}{ |c| }{\backslashbox{$j$}{$i$}} &  1 & 2 & 3 \\ \cline{1-4} %\hline
\multicolumn{1}{ |c| }{1 } &  ${\bm t_2}$ & $t_8$ & $t_8$ \\
\multicolumn{1}{ |c| }{2 } & $t_5$ & $t_6$ & $t_8$\\
\multicolumn{1}{ |c| }{3 } & $t_5$ & $t_8$ & $t_6$ \\
\multicolumn{1}{ |c| }{4 } & $t_8$ & ${\bm t_2}$ & $t_8$ \\
\multicolumn{1}{ |c| }{5 } & $t_6$ & $t_5$  & $t_8$ \\
\multicolumn{1}{ |c| }{6 } & $t_8$ & $t_5$ & $t_6$ \\
\multicolumn{1}{ |c| }{7 } & $t_8$ & $t_8$ & ${\bm t_2}$ \\
\multicolumn{1}{ |c| }{8 } & $t_6$ & $t_8$ & $t_5$\\
\multicolumn{1}{ |c| }{9 } &  $t_8$& $t_6$ & $t_5$ \\
\hline
\end{tabular}
\end{center}
\end{table}
%%%%%%%%%%%%%%%%%%%%%%%%%%%%%%%%%%%%%%%%%%%%%%%%%%%%%%%%%%%%%%%%%%%%%%%%%%%

%%%%%%%%% 6 zero md & 6 zero mds%%%%%%%%%%%%%%%%%%%%%%%%%
\begin{table}[!h]
\caption{Combination of $6$ zero $m_D(i)$ and $6$ zero $m_{DS}(j)$} \label{t5}
\begin{center}
\begin{tabular}{p{2cm}|p{.6cm}p{.6cm}p{.6cm}p{.6cm}p{.6cm}p{.6cm}p{.6cm}p{.6cm}p{.6cm}|}
%\begin{tabular}{c|c|c|}
%\hline\\
\cline{2-10}
 & \multicolumn{9}{c|}{\bf Type of $m_\nu$}\\%{{\bf $8$ zero texture}}\\
\hline
\multicolumn{1}{ |c| }{\backslashbox{$j$}{$i$}} &  1 & 2 & 3 & 4& 5& 6 & 7 & 8 & 9\\ \cline{1-10} %\hline
\multicolumn{1}{ |c| }{1 } &  ${\bm t_1}$  &  ${\bm t_2}$ &  ${\bm t_2}$ & $t_8$ & $t_4$  & $t_8$ & $t_8$ & $t_4$ &$t_8$ \\
\multicolumn{1}{ |c| }{2 } &  ${\bm t_2}$& $t_3$ & $t_5$ & $t_4$ & $t_6$ & $t_6$ & $t_8$ & $t_6$ & $t_7$ \\
\multicolumn{1}{ |c| }{3 } & ${\bm t_2}$ & $t_5$ & $t_3$ &$t_8$ & $t_6$& $t_7$&$t_4$  &$t_6$ &$t_6$ \\
\multicolumn{1}{ |c| }{4 } & $t_8$ &$t_4$  &$t_8$  &${\bm t_1}$ &  ${\bm t_2}$& ${\bm t_2}$ &$t_8$  &$t_8$ &$t_4$ \\
\multicolumn{1}{ |c| }{5 } & $t_4$ &$t_6$  &$t_6$  & ${\bm t_2}$ &$t_3$ &$t_5$ &$t_8$  &$t_7$ &$t_6$ \\
\multicolumn{1}{ |c| }{6 } & $t_8$ & $t_6$ &$t_7$  & ${\bm t_2}$ &$t_5$ &$t_3$ &$t_4$  &$t_6$ &$t_6$ \\
\multicolumn{1}{ |c| }{7 } &  $t_8$&$t_8$  &$t_4$  &$t_8$ &$t_8$ &$t_4$ & ${\bm t_1}$  & ${\bm t_2}$ & ${\bm t_2}$ \\
\multicolumn{1}{ |c| }{8 } & $t_4$ & $t_8$ &$t_6$  &$t_8$ &$t_7$ &$t_6$ &  ${\bm t_2}$ & $t_3$&$t_5$ \\
\multicolumn{1}{ |c| }{9 } &  $t_8$& $t_7$ &$t_6$  &$t_4$ &$t_6$ &$t_6$ &${\bm t_2}$  &$t_5$ & $t_3$\\
\hline
\end{tabular}
\end{center}
\end{table}
%%%%%%%%%%%%%%%%%%%%%%%%%%%%%%%%%%%%%%%%%%%%%%%%%%%%%%%%%%%%%
Table \ref{t2} shows that none of the 9 combinations survives.
Both Table \ref{t3} and \ref{t4} give us 3 viable $m_\nu$s 
where as Table \ref{t5} gives 15 surviving combinations (including both ${\bm t_1}$ and ${\bm t_2}$).
%with combinations ($ij$ value)  $11$, $42$, $73$ for Table \ref{t3}
%and $11$, $24$, $37$, for Table \ref{t4}. 
%\newpage
\section{Parametrization and diagonalization}\label{s5}
In this section at first we write down the explicit forms of the surviving $m_\nu$ s in terms of 
model parameters and again parametrize them in a convenient way. Lets start with Table \ref{t3} considering the 
combination of $6$ zero $m_D$ and $7$ zero $m_{DS}$. For ($i=1,~j=1$)
\begin{equation}
m_\nu=-\frac{1}{m_1} \left(\begin{array}{ccc}  0 & k_1 a_1 y_1 & a_1y_1 \\ 
             k_1 a_1 y_1 & 2kk_1b_1y_1 & (k+k_1)b_1y_1\\
             a_1y_1 & (k+k_1)b_1y_1  & 2b_1y_1 \end{array}\right).\label{m1}
\end{equation}
We assume that the scale factors $k$ and $k_1$ are related by a breaking parameter $\epsilon$ as
$k_1=k(1+\epsilon)$ such that $k$ and $k_1$ becomes equal when $\epsilon$ vanishes.
The above matrix (\ref{m1}) after the substitutions
\begin{equation}
q e^{i \beta}=\frac{a_1y_1}{m_1},~t e^{i\gamma}=\frac{b_1y_1}{m_1}
\end{equation}
becomes
\begin{equation}
m_\nu= \left(\begin{array}{ccc} 0 & kq(1+\epsilon) & q\\ 
             kq(1+\epsilon) & 2k^2te^{i\theta_2}(1+\epsilon) & 2kte^{i\theta_2}+\epsilon kte^{i\theta_2}\\
             q &  2kte^{i\theta_2}+\epsilon kte^{i\theta_2} & 2te^{i\theta_2} \end{array}\right)\label{mf1}
\end{equation}
where we have taken out the phase $\beta$ and the negative sign by the rotation 
\begin{equation} 
m_\nu ~ \rightarrow e^{-\frac{i\pi}{2}}diag(1~e^{-i\beta}~e^{-i\beta}) m_\nu  e^{-\frac{i\pi}{2}} diag(1~e^{-i\beta}~e^{-i\beta})
\end{equation}
and renamed the existing phase as $\theta_2=\gamma-2\beta$. Lets denote the above mass matrix as {\bf Cat I}. 
Effective neutrino mass matrix ($m_\nu$) for other two surviving
combinations, ($i=4,~j=2$) and ($i=7,~j=3$) looks identical to (\ref{mf1}) but with different parametrizations as
\begin{equation}
q e^{i \beta}=\frac{a_2y_2}{m_2},~t e^{i\gamma}=\frac{b_2y_2}{m_2}
\end{equation}
and
\begin{equation}
q e^{i \beta}=\frac{a_3y_3}{m_3},~t e^{i\gamma}=\frac{b_3y_3}{m_3}
\end{equation}
respectively. It is to be noted that we have dubbed the parameters $p,~q,~\beta,~\gamma$ for rest of the cases.\\\\
Moving on to the next set of combinations i.e Table \ref{t4} ($7$ zero $m_D$ and $6$ zero $m_{DS}$), we find that for ($i=1,~j=1$)
\begin{equation}
m_\nu=-\frac{1}{m_1} \left(\begin{array}{ccc}  0 & k b_1 x_1 & b_1x_1 \\ 
             k b_1 x_1 & 2kk_1b_1y_1 & (k+k_1)b_1y_1\\
             b_1x_1 & (k+k_1)b_1y_1  & 2b_1y_1 \end{array}\right)\label{m2}
\end{equation}
which after the combined substitutions
\begin{equation}
 q e^{i \beta}=\frac{b_1x_1}{m_1},~t e^{i\gamma}=\frac{b_1y_1}{m_1}
\end{equation}
and rotation
\begin{equation}
 m_\nu ~ \rightarrow e^{-\frac{i\pi}{2}}diag(1~e^{-i\beta}~e^{-i\beta}) m_\nu  e^{-\frac{i\pi}{2}} diag(1~e^{-i\beta}~e^{-i\beta})
\end{equation}
becomes
\begin{equation}
m_\nu= \left(\begin{array}{ccc} 0 & kq & q\\ 
             kq & 2k^2te^{i\theta_2}(1+\epsilon) & 2kte^{i\theta_2}+\epsilon kte^{i\theta_2}\\
             q &  2kte^{i\theta_2}+\epsilon kte^{i\theta_2} & 2te^{i\theta_2} \end{array}\right)\label{mf2}
\end{equation}
where $\theta_2=\gamma-2\beta$. The resulting neutrino mass matrix (\ref{mf2}) is named as {\bf Cat II}. \\
The $m_\nu$ appeared for other two surviving combinations ($i=2,~j=4$) and 
($i=3,~j=7$) are written in a generic way through ({\ref{mf2}}) with the following 
choices of parameters as 
\begin{eqnarray}
&& qe^{i \beta}=\frac{b_2x_2}{m_2},~t e^{i\gamma}=\frac{b_2y_2}{m_2}\\
&& qe^{i \beta}=\frac{b_3x_3}{m_3},~t e^{i\gamma}=\frac{b_3y_3}{m_3}
\end{eqnarray}
respectively. Both the {\bf Cat I} and {\bf Cat II} matrices have $11$ th element zero indicating the fact that they have 
vanishing $|m_{\nu_{ee}}|$. Unlike the previous two sets of combinations, the third one 
shown in Table \ref{t5} ( $6$ zero $m_D$ and $6$ zero $m_{DS}$ )
possess some combinations for which the resulting $m_\nu$ have all its elements nonzero (denoted by {$\bm t_1$ }).
Table \ref{t5} shows that the combination of $6$ zero $m_D$ and $6$ zero $m_{DS}$ gives rise to a total of $15$ viable structure
of $m_\nu$ s of which $3$ belongs to {$\bm t_1$ } type and remaining $12$ are of {$\bm t_2$ } type. We start with
($i=1,~j=1$) for which explicit form of $m_\nu$ is given by
\begin{equation}
m_\nu=-\frac{1}{m_1} \left(\begin{array}{ccc}  2 a_1x_1 & k b_1 x_1+k_1 a_1y_1 & b_1x_1+a_1y_1 \\ 
             k b_1 x_1+k_1 a_1y_1 & 2kk_1b_1y_1 & (k+k_1)b_1y_1\\
             b_1 x_1+a_1y_1 & (k+k_1)b_1y_1  & 2b_1y_1 \end{array}\right) .
\end{equation}
where $k_1=k(1+\epsilon)$. To get a convenient form of $m_\nu$ we use the parametrizations
\begin{equation}
p e^{i\alpha}=\frac{2 a_1x_1}{m_1},~ qe^{i\beta}=\frac{b_1x_1+a_1y_1}{m_1},~q_1e^{i\beta_1}=\frac{a_1y_1}{m_1},
~te^{i\gamma}=\frac{b_1y_1}{m_1} .
\end{equation}
At first we rewrite the above matrix using these substitutions and then take out the redundant phases using the 
rotation
\begin{equation}
 m_\nu ~ \rightarrow e^{-\frac{i\pi}{2}}diag(e^{-i\frac{\alpha}{2}}~e^{-i\beta}~e^{-i\beta}) m_\nu  e^{-\frac{i\pi}{2}} diag(e^{-i\frac{\alpha}{2}}~e^{-i\beta}~e^{-i\beta}).
\end{equation}
Finally we arrive at a suitable form of $m_\nu$ as
\begin{equation}
m_\nu= \left(\begin{array}{ccc} p & kq+\epsilon k q_1e^{i\theta_1} & q\\ 
             kq+\epsilon k q_1e^{i\theta_1} & 2k^2te^{i\theta_2}(1+\epsilon) & 2kte^{i\theta_2}+\epsilon kte^{i\theta_2}\\
             q &  2kte^{i\theta_2}+\epsilon kte^{i\theta_2} & 2te^{i\theta_2} \end{array}\right)\label{mf3} 
\end{equation}
where the remaining phases are redefined as $\theta_1=\beta_1-\beta$ and $\theta_2=\gamma-2\beta$. This structure of
$m_\nu$ (\ref{mf3}) is designated as {\bf Cat III}.
The other two {$\bm t_1$} type $m_\nu$ s can be brought into this form with some different parametrizations given by
 \begin{equation}
p e^{i\alpha}=\frac{2 a_2x_2}{m_2},~ qe^{i\beta}=\frac{b_2x_2+a_2y_2}{m_2},~q_1e^{i\beta_1}=\frac{a_2y_2}{m_2},
~te^{i\gamma}=\frac{b_2y_2}{m_2}
\end{equation}
for ($i=4,~j=4$) and
\begin{equation}
p e^{i\alpha}=\frac{2 a_3x_3}{m_3},~ qe^{i\beta}=\frac{b_3x_3+a_3y_3}{m_3},~q_1e^{i\beta_1}=\frac{a_3y_3}{m_3},
~te^{i\gamma}=\frac{b_3y_3}{m_3}
\end{equation}
for ($i=7,~j=7$). We can recast all the remaining {$\bm t_2$} type $m_\nu$ to either {\bf Cat I} or {\bf Cat II} and required 
parametrizations are given below. Among these $12$ structures, only 6 are different from each other, i.e we get 6 pairs and
in each pair one is  completely identical to the other. We denote these pairs in a second bracket as
(i)\{($i=1,~j=2$) and ($i=1,~j=3$)\}, (ii)\{($i=4,~j=5$) and ($i=4,~j=6$)\}, (iii)\{($i=7,~j=8$) and ($i=7,~j=9$)\}, 
(iv)\{($i=2,~j=1$) and ($i=3,~j=1$)\}, (v)\{($i=5,~j=4$) and ($i=6,~j=4$)\}, (vi)\{($i=8,~j=7$) and ($i=9,~j=7$)\}.
The first three pairs (i), (ii) and (iii) can be expressed by the generic matrix of {\bf Cat I} with parametrizations
\begin{eqnarray}
&&q e^{i \beta}=\frac{a_1y_1}{m_1},~t e^{i\gamma}=\frac{b_1y_1}{m_1}\nonumber\\
&&q e^{i \beta}=\frac{a_2y_2}{m_2},~t e^{i\gamma}=\frac{b_2y_2}{m_2}\nonumber\\
&&q e^{i \beta}=\frac{a_3y_3}{m_3},~t e^{i\gamma}=\frac{b_3y_3}{m_3}
\end{eqnarray}
respectively, whereas the last three pairs (iv), (v) and (vi) produce that of {\bf Cat II} and the required parametrizations
are
\begin{eqnarray}
&& q e^{i \beta}=\frac{b_1x_1}{m_1},~t e^{i\gamma}=\frac{b_1y_1}{m_1}\nonumber\\
&& q e^{i \beta}=\frac{b_2x_2}{m_2},~t e^{i\gamma}=\frac{b_2y_2}{m_2}\nonumber\\
&& q e^{i \beta}=\frac{b_3x_3}{m_3},~t e^{i\gamma}=\frac{b_3y_3}{m_3}.
\end{eqnarray}
It is clear from the above analysis that all the viable (a total of 21) $m_\nu$ matrices can be written in three categories
namely {\bf Cat I}, {\bf Cat II} and {\bf Cat III} after parametrization. So it is enough to analyze only these three
matrices numerically to examine whether they have any allowed parameter space.  
 
%\begin{eqnarray}

%\end{eqnarray}

\section{Discussion on numerical results}\label{s6}
Now our task is to obtain the exact values of the neutrino oscillation observables (mass squared differences and mixing angles)
of the surviving $m_\nu$ matrices belonging to {\bf Cat I}, {\bf Cat II} and {\bf Cat III}. We use  
straightforward generalized diagonalization methodology developed earlier\cite{Adhikary:2013bma} to calculate mass eigenvalues,
mixing angles and CP violating phases - both Dirac and Majorana type in terms of the mass matrix parameters. 
Neutrino oscillation experimental data generated from global fit (Table \ref{osc}) is
used to obtain the admissible parameter space.
\begin{table}[!ht]
\caption{Input data from neutrino oscillation experiments \label{osc} \cite{Tortola:2012te}}%,\cite{Bennett:2012fp}} 
%\label{input}
\begin{center}
\begin{tabular}{|c|c|}
\hline
{ Quantity} & { $3\sigma$ ranges/other constraint}\\
\hline
$\Delta m_{21}^2$ & $7.12<\Delta m_{21}^2(10^{5}~ eV^{-2})<8.20$\\
$|\Delta m_{31}^2|(N)$ & $2.31<\Delta m_{31}^2(10^{3}~ eV^{-2})<2.74$\\
$|\Delta m_{31}^2|(I)$ & $2.21<\Delta m_{31}^2(10^{3}~ eV^{-2})<2.64$\\
$\theta_{12}$ & $31.30^\circ<\theta_{12}<37.46^\circ$\\
$\theta_{23}$ & $36.86^\circ<\theta_{23}<55.55^\circ$\\
$\theta_{13}$ & $7.49^\circ<\theta_{13}<10.46^\circ$\\
%$\sum_i m_i$ & $<0.44~eV$\\
% $\delta$ & $0-2\pi$\\
\hline
\end{tabular}
\end{center}
\end{table}
%%%%%%%%%%%%%%%%%%%%%%%%%%%%%%%%%%%%%%%%%%%%%%%%%%%%%%%%%%%%
%Before we carry out the detail numerical analysis few important points regarding cosmological bound on sum of three light neutrino
%masses and bound on $|m_{\nu_{ee}}|$ from neutrino less double beta decay experiments have to be mentioned.
%\begin{enumerate}
%\item{A combined analysis of Planck experimental results\cite{Ade:2013zuv} with different other cosmological experiments 
%suggests that the upper bound of sum of three neutrino masses may vary inside a range as 
%$\sum m_i(=m_1+m_2+m_3)$ $<$ $(0.23 -1.11)eV$ \cite{Giusarma:2013pmn} . The upper value of the above range arises from a set of 
%experimental results comprises with Planck, WMAP low l polarization\cite{Bennett:2012zja}, gravitational lensing and results prior on
%the Hubble constant $H_0$ from the Hubble space telescope data whereas the lower value obtained due to the 
%inclusion of SDSS DR8\cite{Aihara:2011sj} result with the above combination. 
%} 
%\item{Neutrinoless double beta decay experiment\cite{Tortola:2012te,Giuliani:2010zz,Rodejohann:2012xd} constrains the neutrino mass matrix element $|m_{\nu_{ee}}|$.
%EXO-200 experiment\cite{Auger:2012ar} has given a range of the 
%upper limit of $|m_{\nu_{ee}}|$ as $|m_{\nu_{ee}}|$ $<$ $(0.14-0.38)eV$. 
%}
%\end{enumerate}
%%%%%%%%%%%%%%%%%%%%%%%%%%%%%%%%%%%%%%%%%%%%%%%%%%%%%%%%%%%%%%%%%%%%%%%%%%%
In this work the experimental constraints used to restrict the parameters are solar and atmospheric mass squared differences 
and three mixing angles and we predict the individual neutrino masses, the corresponding hierarchy, their sum ($\sum m_i$), 
the value of $|m_{\nu_{ee}}|$, the CP violating Jarlskog invariant $J_{CP}$ and the Dirac CP violating phase $\delta_D$.
%We also predict the value of the Majorana phases, however, at present their testability is far away. 
We also predict the value of the Majorana phases, which will be tested \cite{Bahcall:2004ip,Cremonesi:2012av} 
in neutrinoless double beta decay experiments, however, determination of their values is a challenging task.
\paragraph{}
First we analyze the {\bf Cat I} and {\bf Cat II} matrices where we encounter the vanishing $|m_{\nu_{ee}}|$ element. Although
the explicit structure of these two matrices are different from each other they are composed of same $5$ parameters namely
$k,~q,~t,~\epsilon$ and a phase parameter $\theta_2$. After scanning those parameters in various possible ranges we find that
both of them ({\bf Cat I} and {\bf Cat II}) fail to produce all the neutrino oscillation observables simultaneously inside
the allowed $3 \sigma$ range as mentioned above (Table \ref{osc}). It has been observed that both of
the above matrices can produce all the experimental observables except $\theta_{13}$ inside the allowed range.
The lowest $\theta_{13}$ produced here exceeds the upper limit of the $3\sigma$ range quoted in Table \ref{osc}.  
Hence, the $m_\nu$ matrices grouped in {\bf Cat I} and {\bf Cat II} are discarded. We are now left with only one type of
$m_\nu$ ({\bf Cat III}). 
%We have to examine whether this $m_\nu$ having all its elements nonzero can produce the mixing
%angles and mass eigenvalues within the experimentally obtained $3\sigma$ limit. 
The matrix belonging to  {\bf Cat III}
is made up of total $8$ parameters and they are $p,~q,~k,~t,~q_1,~\epsilon$ and two phase parameters $\theta_1$, $\theta_2$.
Varying those parameters in nearly all possible ranges we find some admissible parameter space satisfying extant data.
Ranges of the allowed parameters for which the values of the resulting oscillation observables fall within $3\sigma$ range
of extant data are shown in the Table \ref{range} below.
\begin{table}[!ht]
\caption{Allowed ranges of parameters \label{range} }%,\cite{Bennett:2012fp}} 
%\label{input}
\begin{center}
\begin{tabular}{|c|c|c|c|c|c|c|}
\hline
 {\bf Parameters} & $p$ & $q$ & $k$ & $t$& $q_1$ & $\epsilon$ \\%& $\theta_1$ (deg.) & $\theta_2$ (deg.)\\
\hline
{\bf Allowed} & $0.001$-$0.016$&$0.001$-$0.053$ &$0.2$-$0.9$ &$0.006$-$0.028$&$0.01$-$0.1$&$1.4$-$9.5$ \\%& &\\
{\bf ranges}  & & & & & & \\%& &\\
\hline
\end{tabular}
\end{center}
\end{table}
%The allowed phase parameter space consists of four different patches in $\theta_1$ vs $\theta_2$ plane. Lets denote
%them as patch1, patch2, patch3, patch4 where patch4 is mirror image to patch1 and patch3 is mirror image to patch2.
%Ranges of the two phases in patch1 and patch2 are listed below in Table \ref{phase}.
The phase parameter space is divided in four patches in $\theta_1$ vs $\theta_2$ plane and pairwise one is mirror image to
the other. The allowed values of $\theta_1$ and $\theta_2$ are shown in Table \ref{phase}.
%%%%%%%%%%%%%%%%%%%%%%%%%%%%%%%%%%%%%%%%%%%%%%%%%%%%%%%%%%%%%%%%%%%%%%%%%%%%%%%%%%%%%%%%%%
%\begin{table}[!h]
%\caption{Allowed phase parameter space in $\theta_1$ vs $\theta_2$ plane} \label{phase}
%\begin{center}
%\begin{tabular}{c|c|c|}
%\cline{2-3}
%& $\theta_1$ (deg.)& $ \theta_2$ (deg.) \\%\multicolumn{2}{c|}{\bf Type of $m_\nu$}\\%{{\bf $8$ zero texture}}\\
%\hline
%\multicolumn{1}{ |c| }{{\bf Patch1}} &  $(-180)$-$(-73.2)$ &$(-35)$-$(-180)$  \\ \cline{1-3} %\hline
%\multicolumn{1}{ |c| }{{\bf Patch2} } & $(-112.3)$-$(-35.5)$&$(-180)$-$(-54.3)$ \\
%\hline
%\end{tabular}
%\end{center}
%\end{table}
%%%%%%%%%%%%%%%%%%%%%%%%%%%%%%%%%%%%%%%%%%%%%%%%%%%%%%%%%%%%%%%
\begin{table}[!h]
\caption{Allowed ranges of $\theta_1$ and $\theta_2$ phase parameters} \label{phase}
\begin{center}
\begin{tabular}{|c|c|}
\hline
 $\theta_1$ (deg.)& $ \theta_2$ (deg.) \\
\hline
 $(-180)$-$(-73.2)$ &$(-35)$-$(-180)$  \\  
\hline
\multicolumn{2}{|c|}{and}\\
\hline
 $(-112.3)$-$(-35.5)$&$(-180)$-$(-54.3)$ \\
\hline
\end{tabular}
\end{center}
\end{table}
In Table \ref{range1} we predict the individual mass eigenvalues and the sum of the three neutrino masses ($\sum m_i$)
and the value of $|m_{\nu_{ee}}|$. 
\paragraph{}
\begin{figure}
\includegraphics[width=5.5cm,height=5.5cm,angle=0]{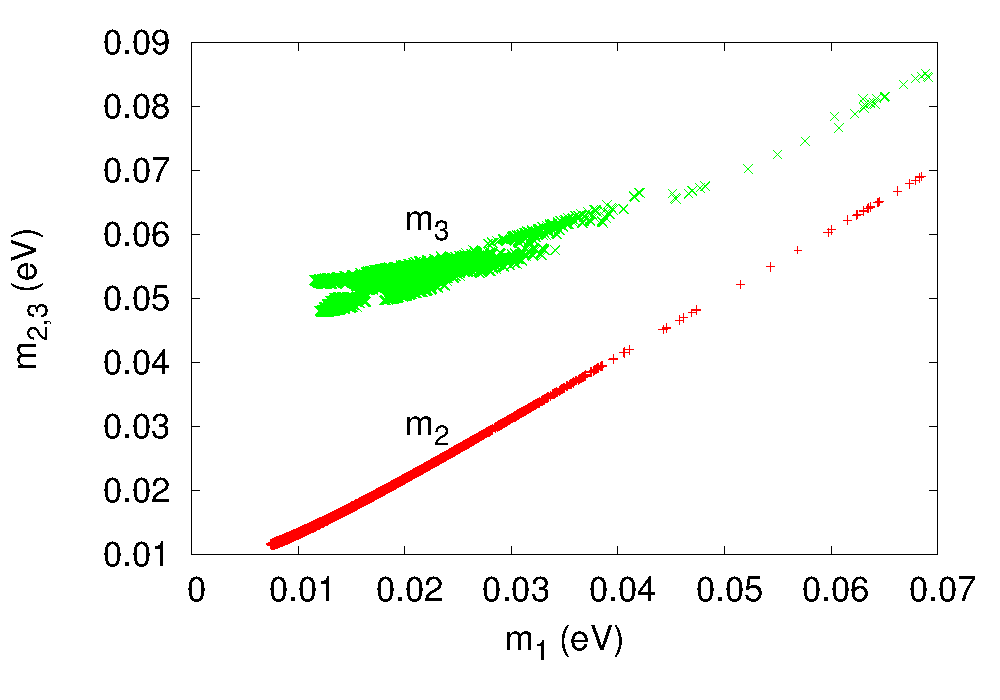} 
\includegraphics[width=5.5cm,height=5.5cm,angle=0]{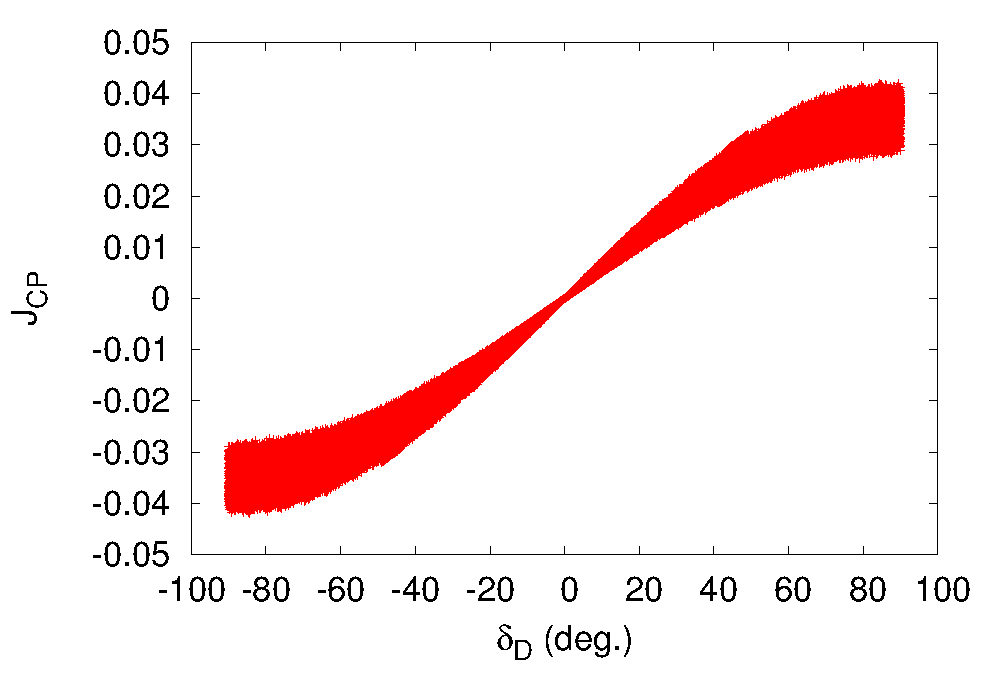}
\includegraphics[width=5.5cm,height=5.5cm,angle=0]{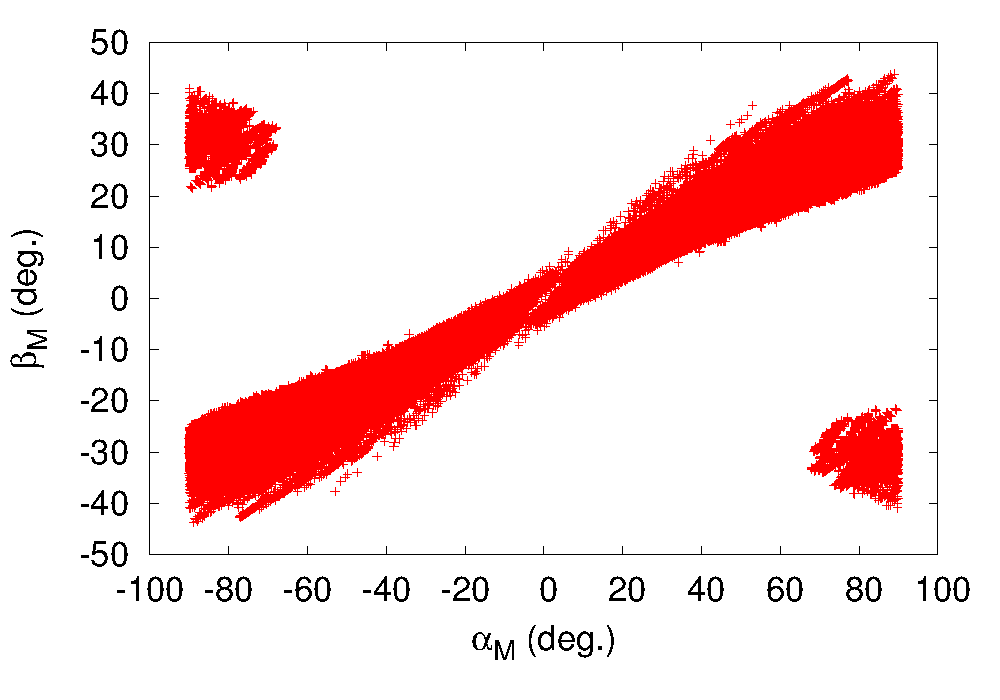}
\caption{(colour online) Plot of mass eigenvalues $m_1$ vs $m_{2,3}$ (left), 
Jarlskog measure($J_{CP}$) vs Dirac CP Phase($\delta_D$) (middle) and 
Majorana Phases ($\alpha_M$ vs $\beta_M$) (right)}
\label{f1}
\end{figure}
Some comments on the issue of the predictions of the present scheme are in order.
\begin{enumerate}
\item {First of all the mass ordering obtained in the present scheme is normal and it is illustrated in 
the left panel of Fig.\ref{f1} through 
a plot of $m_1$ with $m_2$ and $m_3$. It has been shown\cite{Prakash:2013nra,Agarwalla:2013ju,Chatterjee:2013qus} 
that precise determination of $\theta_{13}$ through 
reactor neutrino experiments will enable us to fix the neutrino mass ordering through a combined analysis complying 
with the results of long baseline experiments NO$\nu$A\cite{Ayres:2004js,noa} and T2K\cite{Abe:2011sj}, since, result of only one of them is insufficient to
probe the mass ordering due to degenerate nature of $\delta_{CP}$ in the expression of P$(\nu_\mu \rightarrow \nu_e)$\cite{Prakash:2013nra,Prakash:2013dua}.
Thus the prediction of the hierarchy of the present scheme will be tested in near future.
\paragraph{}
Next we have plotted $J_{CP}$ with $\delta_D$ in the middle panel of Fig.\ref{f1}. Information on the value of 
$J_{CP}$ can be obtained from the experiment looking for the difference between  P$(\nu_\mu \rightarrow \nu_e)$
and P$(\overline{\nu_\mu} \rightarrow \overline{\nu_e})$ using neutrino and antineutrino beams. A detailed
review on this issue is given in Ref.\cite{Minakata:2008yz}}.
%%%%%%%%%%%%%%%%%%%%%%%%%%%%%%%%%%%%%%%%%
%Next, the above analysis will not only probe the hierarchy of neutrino masses but, as well as, The CP violating Jarlskog
%measure $J_{CP}$ and $\delta_D$ will also be determined. In the middle panel of Fig.\ref{f1} we show the variation of $J_{CP}$ with $\delta_D$
%arises in the present model and their allowed range is given in Table \ref{r3}}.
%%%%%%%%%%%%%%%%%%%%%%%%%%%%%%%%%%%%%%%%%%%%
\item{The sum of three neutrino masses ($\sum m_i$) is always below the present cosmological experimental bound 
($\sum m_i<0.23eV $)\cite{Ade:2013zuv,Giusarma:2013pmn,Bennett:2012zja,Aihara:2011sj}. However, 
the next analysis\cite{Lesgourgues:2014zoa} of Planck CMB satellite data in combination
with more sensitive other cosmological and astrophysical experiments, such as Baryon oscillation spectroscopic survey(BOSS),
The Dark energy survey(DES), The Large synoptic survey telescope(LSST) and the Euclid satellite, will bring down the lower
limit in a region of $m_\nu\sim0.1eV$ for inverted ordering and for normal mass ordering of neutrinos it will be pushed down to
$m_\nu\sim0.05eV$. Thus most of the present predicted range of $\sum m_i$ $(\sim0.068-0.22)eV$ will be under scanner in
the near future.
}  
\item{In the present work, the matrix element $|m_{\nu_{ee}}|$, which is constrained by the neutrinoless double beta decay
($\beta\beta0\nu$) experiment\cite{Giuliani:2010zz,Rodejohann:2012xd,Bilenky:2014uka} varies within a range as shown in Table \ref{range1}. EXO-200 experiment\cite{Auger:2012ar}
has given a range on the upper limit of $|m_{\nu_{ee}}|$ as $|m_{\nu_{ee}}|<(0.14-0.38)eV$. Thus the predicted
values of the present work are below the above experimental value and are beyond the reach to be testified. However, it has been
claimed that NEXT-100 experiment\cite{DavidLorcafortheNEXT:2014fga} will probe the value of $|m_{\nu_{ee}}|\sim 0.1eV$. We go optimistically with 
such findings in the near future.
\paragraph{}
We also provide a plot of Majorana phases in the right panel of Fig.\ref{f1} and their allowed range is presented in Table \ref{r3}.
Determination of Majorana nature of neutrinos requires
positive evidence from $\beta\beta0\nu$ experiment. However, in such process, CP symmetry is conserved, and, hence, to
probe CP violating Majorana phases one has to look for the Lepton Flavor violating processes also\cite{Xing:2013woa}.
}
\end{enumerate}
\begin{table}[!ht]
\caption{Allowed values of individual neutrino masses ($m_i$) and their sum ($\sum m_i$) and $|m_{\nu_{ee}}|$  \label{range1} }%,\cite{Bennett:2012fp}} 
%\label{input}
\begin{center}
\begin{tabular}{|c|c|c|c|c|}
\hline
 $m_1$ (eV)& $m_2$ (eV) & $m_3$ (eV) &$\sum m_i
$ (eV) & $|m_{\nu_{ee}}|$ (eV)\\
\hline
$0.007$-$0.068$ &$0.011$-$0.069$ &$0.047$-$0.085$&$0.068$-$0.22$ &$0.001$-$0.016$ \\
 & & & & \\
\hline
\end{tabular}
\end{center}
\end{table}
%\newpage
\begin{table}[!ht]
\caption{Allowed values of Jarlskog measure ($J_{CP}$), Dirac CP Phase ($\delta_D$) and Majorana Phases ($\alpha_M$ and $\beta_M$)  \label{r3} }%,\cite{Bennett:2012fp}} 
%\label{input}
\begin{center}
\begin{tabular}{|c|c|c|c|}
\hline
 $J_{CP}$& $\delta_D$(deg.) & $\alpha_M$(deg.)  & $\beta_M$(deg.) \\
\hline
 $(-0.041)$-$(0.041)$&$(-90)$-$(90)$ &$(-90)$-$(90)$ &$(-41)$-$(41)$ \\
 & & & \\
\hline
\end{tabular}
\end{center}
\end{table}
\section{Summary and conclusion}\label{s7}
Our goal of this work is to describe a phenomenologically viable effective light neutrino mass matrix ($m_\nu$) with
minimum number of parameters. The light neutrino mass matrix $m_\nu$ is generated through linear seesaw mechanism
where along with standard $SU(2)_L\times U(1)_Y$ particle contents, three right chiral singlet neutrinos ($N_{R_i}$) and  
three other fermion singlets ($S_{R_i}$) are present. The $9\times 9$ Majorana mass matrix obtained in this basis is
further block diagonalized to get mass matrix for the light neutrinos. After imposing the assumption of absence
of global $U(1)_L$ symmetry breaking term we get the final working formula for $m_\nu$ which is composed of three
matrices $m_D$, $m_{DS}$ and $m_{RS}$. Without sacrificing any generality we are allowed to choose the charge lepton
mass matrix and $m_{RS}$ to be diagonal. To reduce the number of independent parameters our next idea is to invoke
scaling ansatz in the two Dirac type matrices $m_D$ and $m_{DS}$. The scaling ansatz is broken in final $m_\nu$
by choice of different scale factors for $m_D$ and $m_{DS}$ to get rid of vanishing value of $\theta_{13}$.
\paragraph{}
The most important part of our analysis is to accommodate as many zeros as possible in those scaling ansatz invariant
$m_{D}$ and $m_{DS}$. It is noticed that we can get at most three $7$ zero textures and nine $6$ zero textures for both 
$m_D$ and $m_{DS}$. So their combination give rise to $12\times 12~=144$ $m_\nu$ matrices. Depending upon the positions
of zeros and nonzero elements these $144$ textures are generically represented by eight matrices denoted as 
$t_i$ ($i=1,2...8$) of which $t_1$ and $t_2$ are phenomenologically viable and the rest six are discarded. Among
$144$ $m_\nu$ matrices we get eighteen $t_2$ type matrices and three $t_1$ type matrices and the notable fact is that all three
$t_1$ type matrices are generated by the combination of $6$ zero $m_D$ and $6$ zero $m_{DS}$ where as $t_2$ type 
matrices emerged in all type of combinations except those of $7$ zero $m_D$ with $7$ zero $m_{DS}$.
\paragraph{}
All the $t_2$ type matrices are recasted in two types of $m_\nu$  ({\bf Cat I} and {\bf Cat II}) and $t_1$ type matrices
can be represented by single $m_\nu$ matrix ({\bf Cat III}) after phase redefinition and reparametrization. The numerical analysis is done
thereafter. It is clear from the detailed numerical analysis that {\bf Cat I} and {\bf Cat II} matrices are disfavoured
by oscillation data and the only surviving $m_\nu$ belongs to {\bf Cat III}. The mass ordering of the light neutrinos is
normal and the value of $\sum m_i$ is also below the present experimental lower limit. We conclude with a comment that to meet the 
phenomenological requirements we need at most two $6$ zero matrices ($m_D$ and $m_{DS}$) and one diagonal matrix $m_{RS}$ 
while working with linear seesaw mechanism. Increase in number of zeros in any of the two Dirac type matrices will make 
the resulting $m_\nu$ phenomenologically invalid. Our numerical analysis of the survived texture, predicts quantitative
nature of neutrino mass hierarchy and other observables, among them, except Majorana phases, all of them will be probed in 
the near future.
\\\\\\
%\newpage
%\vspace{1cm}
{\bf Acknowledgment}\\
M.C. and A.G are thankful to  B. Adhikary for helpful discussions. 
H.Z.D. acknowledges the Saha Institute of Nuclear Physics for its hospitality while this work is in progress.\\
%HZD thanks the Saha Institute of Nuclear Physics for its hospitality while this work is in progress.\\
%\vspace{1cm}

%

\end{document}